\documentclass[twocolumn,showpacs,preprintnumbers,amsmath]{revtex4}
\usepackage{amssymb}

\usepackage{graphicx}
\usepackage{dcolumn}
\usepackage{bm}

\begin{document}

\title{Vortex-lattice melting in magnesium diboride in terms of the elastic theory}
\author{Qing-Miao Nie$^{1,2}$, Jian-Ping Lv$^{2,3}$, and Qing-Hu Chen$^{2,3,*}$ }

\address{
$^{1}$ Department of Applied Physics, Zhejiang University of
Technology, Hangzhou 310023, P. R. China \\
$^{2}$ Center for Statistical and Theoretical Condensed Matter
Physics, Zhejiang Normal University, Jinhua 321004, P. R. China \\
$^3$ Department of Physics, Zhejiang University, Hangzhou 310027, P.
R. China}

\date{\today}

\begin{abstract}
In the framework of elastic theory, we study the vortex-lattice
melting transitions in magnesium diboride for magnetic fields both
parallel and perpendicular to the anisotropy axis. Using the
parameters from experiments, the vortex-lattice melting lines in the
H-T diagram are located systematically by various groups of
Lindemann numbers and the anisotropic parameters. It is observed
that the theoretical result for the vortex melting with parallel and
perpendicular fields agrees well with the  experimental data.
\end{abstract}


\maketitle

\section{\label{sec:level1}Introduction}

Since the discovery of superconductivity with $T_c= 39K$ in
magnesium diboride (MgB2) was announced \cite{Nagamatsu}, it has
caused  a large number of experimental
\cite{Bud¡¯ko,Kotegawa,Kortus,Liu,Bouquet,Wen,Eltsev} and
theoretical investigations \cite{Angst, Askerzade}. This
introduced a new, simple (three atoms per unit cell) binary
intermetallic superconductor with a record high (by nearly a
factor of 2) superconducting transition temperature for a nonoxide
and non-C60-based compound. It displays a variety of unusual
properties. The high transition temperature seems to be either
above or at the limit suggested theoretically several decades ago
for BCS, phonon-mediated superconductivity \cite{McMillan}. It is
characterized by a double band superconductor with two
superconducting gaps of different size, the larger one originating
from a quasi-2D $\sigma$-band and the smaller one from a 3D
$\pi$-band. The electronic $\sigma$-states are confined to the
boron planes and couple strongly to the in-plane vibration of the
boron atoms. The unfixed anisotropic parameter $\gamma$ vary
widely , ranging from 1.1 to 6 depending on the measurement
technique and on sample types\cite{Angst,Pradhan,Sologubenko,Kim}.
Although, some experimental data  can shed light on the mechanism
of superconductivity, it was demonstrated \cite{Koshelev} that the
type-II superconductivity may not be well  described by the
standard anisotropic Ginzburg-Landau theory, just due to the
two-band structure.

The vortex-lattice solid (glass with random pinning) state with zero
linear resistivity is crucial for the application of high-$T_{c}$
superconductors, thus the melting of vortex-lattice in bulk type-II
superconductors is of great
significance~\cite{brandt,Houghton,Safar,cubitt,Herbut}. The
traditional Lindemann theory suggests that the lattice melts when
the root mean square thermal displacements of the components of a
lattice reach a certain fraction of the equilibrium lattice spacing.
This criterion was first adopted to study the vortex-lattice melting
transition in type-II superconductors with a magnetic field parallel
to the anisotropic axis\cite{Houghton}, then it was used to draw the
melting lines when magnetic field is perpendicular to the
anisotropic axis~\cite{blatterG,Calson,huchen,nie}. Since the upper
critical field is also high in $MgB_2$, the thermal fluctuation may
drive the vortex-lattice to a vortex liquid in a field far below the
upper critical field \cite{blatterG} through vortex melting.

In this paper, using the parameters measured in recent two typical
experiments\cite{Wen,Eltsev}, we study the vortex-lattice melting
transitions in the framework of the elastic theory
phenomenologically. The melting lines for  magnetic fields both
parallel and perpendicular to the anisotropic axis are
systematically located with different groups of Lindemann numbers
and anisotropic parameter. The comparison with experiments are
also made. The present paper is organized as follows.  In Section
2, we introduce the theoretical method. Section 3 presents the
main results. We give a short summary in the final section.

\section{Elastic theory}

For ideal triangular vortex-line lattice,  the free energy in
elastic theory for fields  parallel and perpendicular to the x axis
can be generally written  with quadratic terms of the deviation
vector $\mathbf{u}=(u_x,u_y)$  describing the fluctuations of
vortices from their equilibrium
positions\cite{brandt,Houghton,blatterG,Calson,huchen,nie,chen1}
\begin{equation}
 F=\frac 12\int\limits_{}\frac{d^3\mathbf{k}}{(2\pi
)^3}\mathbf{u}\cdot \mathbf{C}\cdot \mathbf{u},
\end{equation}
The matrix $C$ for  parallel fields    is different from that for
perpendicular fields. We denote $\mathbf{C^c}$ and $\mathbf{C^{ab}}$
to be the elastic matrix for the fields parallel and  perpendicular
to c-axis, respectively, which are given as follows
\begin{equation}
\mathbf{{C^c}=\left(
\begin{array}{cc}
c_Lk_x^2+c_{66}k_{\perp }^2+c_{44}k_z^2 & c_Lk_xk_y \\
c_Lk_xk_y & c_Lk_y^2+c_{66}k_{\perp }^2+c_{44}k_z^2
\end{array}
\right) }
\end{equation}
and
\begin{equation}  \label{e.2}
\mathbf{{C^{ab}}=\left(
\begin{array}{cc}
c_{11}k_{x}^{2}+c_{66}^{h}k_{y}^{2}+c_{44}^{h}k_{z}^{2} &
c_{11}k_{x}k_{y}
\\
c_{11}k_{x}k_{y} &
c_{66}^{e}k_{x}^{2}+c_{11}k_{y}^{2}+c_{44}^{e}k_{z}^{2}
\end{array}
\right)}
\end{equation}
where $k_{\perp }^2=k_x^2+k_y^2$, $c_{66}, c_L=c_{11}-c_{66}$, and
$c_{44}$ are the wave-vector-dependent shear, bulk, and tilt elastic
moduli. The detailed expressions for these  moduli can be found in
several previous papers \cite{brandt,Houghton,Calson,nie,chen1}. The
thermal fluctuations of the vortices are given by inverting the
kernel $  \mathbf{C(k)}$
\begin{eqnarray}
\langle u_\alpha ^2\rangle=\frac{k_BT}{(2\pi )^3}\int
d\mathbf{k}[\mathbf{C}^{\beta}_{\alpha\alpha}]^{-1}(\mathbf{k}),
\alpha =x,y,  \beta =c,ab.
\end{eqnarray}

The Lindemann criterion presumes that the lattice melts, when the
root mean square thermal displacements of the components of a
lattice reach some fraction of the equilibrium lattice spacing. For
parallel fields,  we consider the mean-square displacement of a
vortex lattice  from the equilibrium $ d^2(T)=u_x^2+u_y^2$, then use
the usual isotropic Lindemann criterion ,
\begin{equation}
\label{e.26} \langle d^{2}\rangle =c^{2}a^{2}.
\end{equation}
For perpendicular fields, an  anisotropic Lindemann criterion should
be  employed
\begin{equation}
\label{e.26} \langle u_{x}^{2}\rangle =c_{x}^{2}a_x^{2},{\;\;\;}
\langle u_{y}^{2}\rangle =c_{y}^{2}a_y^{2}.
\end{equation}
with $c_x$ and $c_y$ are two Lindemann numbers for two transverse
directions. Combining the Lindemann criterion and the elastic
theory, the  melting equations are derived  with the  Lindemann
coefficients $c's$.

If fields are applied perpendicular to c-axis, the effect of layer
pinning reduces fluctuations in both directions and induces an
additional momentum-independent term to the elastic matrix in Eq.
(3) such that
\begin{equation}
\label{e.25} {\bf C^{ab}_{lp}}=\left(
\begin{array}{cc}
c_{11}k_{x}^{2}+c_{66}^{h}k_{y}^{2}+c_{44}^{h}k_{z}^{2}+\Theta &
c_{11}k_{x}k_{y} \\
c_{11}k_{x}k_{y} &
c_{11}k_{y}^{2}+c_{66}^{e}k_{x}^{2}+c_{44}^{e}k_{z}^{2}
\end{array}
\right),
\end{equation}
where $\Theta$ is proportional to the critical depinning current and
depends on the layer separation $s$~\cite{Ivlev} .

\section{Results and discussions}

\subsection{Vortex melting lines for fields parallel to the c-axis}

\begin{figure}[tu1]
\begin{center}
\includegraphics[scale=0.8]{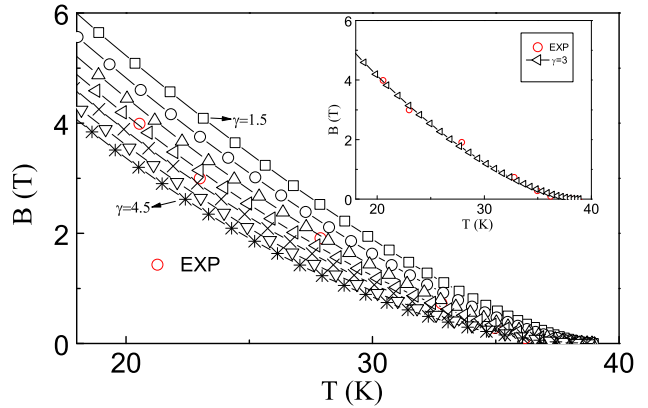}
\vspace{-0.8cm} \caption{\label{whh} The comparison of
vortex-lattice melting lines for magnetic fields parallel to c-axis
with experimental results in Ref. ~\cite{Wen}. The anisotropic
parameter $\gamma$ varies from 1.5 to 4.5 with the step 0.5( from
top to bottom ). The inset shows the melting line with the
anisotropic parameter $\gamma=3$ for Lindemann number $c_x=0.1$.}
\end{center}
\end{figure}

First, we  calculate  the melting line of $MgB_2$ thin film  for
fields parallel to the c-axis. The experiments were performed by Wen
et al. \cite{Wen}  We employ the experimental parameters as follows:
the zero-field superconducting transition temperature $T_{c}=39$K,
The Ginzburg-Landau parameters $\kappa=26$, the upper critical
fields along $c$ axis $H_{c2}^{c}(T=0)=15T$. Since the anisotropy
coefficient $\gamma = H_{c2}^{ab}/H_{c2}^c$, is not well established
for $MgB_2$\cite{Angst,Pradhan,Sologubenko,Kim}. We vary the
anisotropic parameter $\gamma$ in the range of 1.5 to 4.5 in our
calculation. When fields parallel to the c-axis, we set $c_x=c_y$
for no anisotropy exists here. We draw the melting lines by elastic
theory with the Lindemann number $c_x=c_y=0.1$ and with the
variation of $\gamma$, as shown in Fig.1. Interesting, in the inset
of Fig.1 we find that the melting line with $\gamma=3$ agrees well
with the irreversibility lines measured by SQUID in ref. \cite{Wen}.
The irreversibility line in superconductor is usually regarded as
the melting line. \cite{blatterG}. The value of $\gamma=3$ in this
sample is similar to that on transport measurements of the upper
critical field anisotropy performed on single crystals
\cite{Pradhan,Kim}.

\begin{figure}[tu2]
\begin{center}
\includegraphics[scale=0.7]{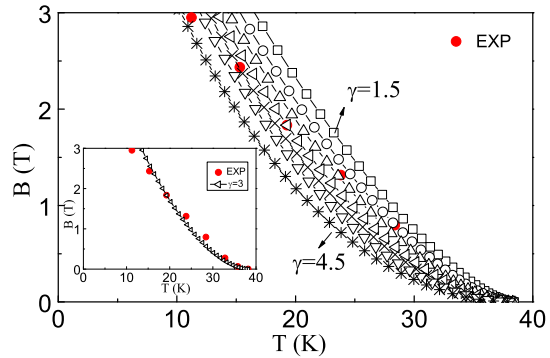}

\vspace{-0.3cm}\caption{\label{Figure Graph1} The vortex-lattice
melting lines of $MgB_2$ single crystal for magnetic fields
parallel to c-axis. The Lindemann number $c_x=0.1$ and the
anisotropic parameter $\gamma$ varies from 1.5 to 4.5 with the
step 0.5( from top to bottom ). Solid circles are experimental
results from Ref. ~\cite{Eltsev}. The inset shows the comparison
of melting line with $\gamma=3$ and $c_x=0.1$ with the
experimental results ~\cite{Eltsev}. }
\end{center}
\end{figure}

Then, we evaluate the  melting line of  $MgB_2$ single crystal for
fields parallel to the c-axis. This experiment was done by Eltsev
et al. \cite{Eltsev}  The  experimental parameters measured from
Ref. \cite{Eltsev} are collected  as follows: $T_{c}=38.5$K,
$\kappa=26$, $H_{c2}^{c}(T=0)=7.5T$. We also change the
anisotropic parameter $\gamma$ from $1.5$ to $4.5$ as above. It is
found that the variation of $\gamma$  also has influence on the
melting line. Fig. 2 shows the theoretical melting results with
$c_x=c_y=0.1$, where the experimental data are also exhibited. As
shown in the inset of Fig. 2, we observe that the melting line
with $\gamma=3$ agrees with experimental data \cite{Eltsev}. Here
the resulted anisotropic parameter $\gamma=3$ also matches up to
the experiment results on transport measurements of  $MgB_2$
single crystals \cite{Pradhan,Kim}.

\subsection{Vortex melting lines for fields perpendicular to the c-axis}

We next turn to  the vortex melting transition for fields
perpendicular to c-axis. The experimental parameters for a
$MgB_2$ single crystal are taken from Ref. \cite{Eltsev} as the
following: $T_{c}=38.5$K, $\gamma=3$, $\kappa=26$,
$H_{c2}^{ab}(T=0)=22T$. Following the procedure outlined in the
preceding section, we calculate the thermal fluctuations in the
two transverse directions. Setting $C_{x}=C_{y}=0.14$, we obtain
two curves similar to those in Ref. \cite{Calson}, which are shown
in Fig. 3.   As pointed out in Ref.~\cite{huchen}, to interpret
these two curves as two " melting lines " and thus reach to a
conclusion of an intermediate smectic phase is unphysical, since
the elastic theory can give at best one single melting line.  To
impose the same Lindemann number along the two directions is of no
physical basis. In order to achieve a single melting line, we tune
the Lindemann number $c_y$ at 0.1. A good collapse of the melting
lines in two directions can be achieved if setting the ratio $
c_x/c_y \approx 1.4$, as shown in Fig. 3. It is interesting to
note that this ratio is very close to $1.37$ observed in Ref.
\cite{nie} using parameters in cuprate superconductors. More
importantly, we  reproduce reasonably the experimental melting
line in Ref. \cite{Eltsev} as shown in fig. 3.

The vortex melting for fields perpendicular to the c-axis is also
influenced by the layer pinning. In order to study this intrinsic
layer pinning  effect,  the matrix (7) is used to calculate the
thermal fluctuations along two transverse directions. Here, the
layer separation of magnesium diboride is $s=3.524 \AA
 $\cite{Nagamatsu}. The melting lines for $\gamma=3$ with Lindemann number $c_{x}=c_{y}=0.12$
are collected in Fig. 4. We find that the melting line with
$c_{x}=0.12$ matches the experimental data. By tuning the
Lindemann number $c_y$, a good collapse can also be obtained by
setting the same ration $ c_x/c_y \approx 1.33$.

\begin{figure}[tu3]
\begin{center}
\includegraphics[scale=0.7]{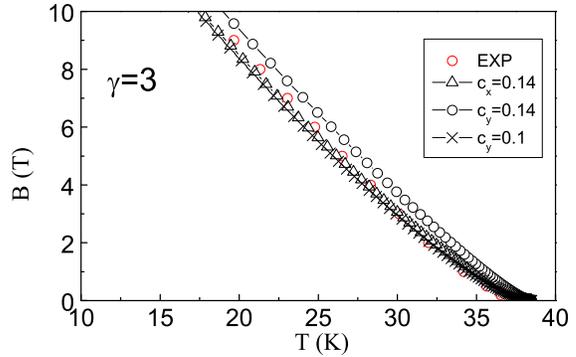}

\vspace{-0.8cm}\caption{\label{Figure Graph1} The comparison of
vortex-lattice melting lines for magnetic fields perpendicular to
c-axis with experimental results in Ref. ~\cite{Eltsev} (without
intrinsic layer pinning).}
\end{center}
\end{figure}

\begin{figure}[tu4]
\begin{center}
\includegraphics[scale=0.7]{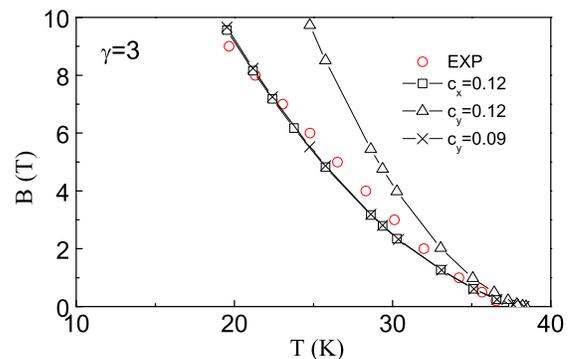}

\vspace{-0.8cm}\caption{\label{Figure Graph1}  The comparison of
vortex-lattice melting lines for magnetic fields perpendicular to
c-axis with experimental results in Ref. ~\cite{Eltsev} (with
intrinsic layer pinning).}
\end{center}
\end{figure}

\section{Summary}

In the framework of the elastic theory, we have studied the melting
transition in magnesium diboride for magnetic fields parallel and
perpendicular to the anisotropy axis. Using the parameters from
experiments, the melting lines with various Lindemann numbers and
the anisotropic parameter are composed. Although the anisotropic
parameter varies with the different measurement technique and sample
types, the melting lines derived here  agree  well with those in
thin films \cite{Wen} and single crystal\cite{Eltsev} by using
Lindemann number $c_x=0.1$ and anisotropic parameter $\gamma=3$ for
magnetic field parallel to the $c-$axis. For fields perpendicular to
the c-axis, it is observed that thermal fluctuations normalized by
vortex separations in the two transverse directions are proportional
to each other, similar to those observed in cuprate superconductors.
The ratio $ c_x/c_y$ to achieve  a single melting line is very close
to that observed in Ref. \cite{nie}.  More interestingly, by using
$c_x=0.12$ and $c_y=0.09$, we are able to draw the melting line
which fits the experimental melting line in ~\cite{Eltsev} quite
well. Although the standard anisotropic Ginzburg-Landau theory may
not be applicable to magnesium diboride, the elastic theory of
vortex matter can provide a good description of the vortex melting
in this two-band superconductors.

\begin{acknowledgments}

This work was supported by National Natural Science Foundation of
China under Grant Nos. 10774128(QHC)  and 10804098(QMN), PCSIRT
(Grant No. IRT0754) in University in China,  National Basic Research
Program of China (Grant Nos. 2006CB601003 and 2009CB929104),
Zhejiang Provincial Natural Science Foundation under Grant No.
Z7080203, and Program for Innovative Research  Team  in Zhejiang
Normal University,
\end{acknowledgments}

$^{*}$ qhchen@zju.edu.cn

\end{document}